\documentclass[conference]{IEEEtran}
\IEEEoverridecommandlockouts
\usepackage{cite}
\usepackage{amsmath,amssymb,amsfonts}
\usepackage{graphicx}
\usepackage{siunitx}
\usepackage{textcomp}
\usepackage{xcolor}
\usepackage{csquotes}
\usepackage{algorithm,algorithmicx}
\usepackage[noend]{algpseudocode}

\def\BibTeX{{\rm B\kern-.05em{\sc i\kern-.025em b}\kern-.08em
    T\kern-.1667em\lower.7ex\hbox{E}\kern-.125emX}}

\begin{document}

\title{Interaction-Aware Sensitivity Analysis for Aerodynamic Optimization Results using\\Information Theory}

\author{\IEEEauthorblockN{1\textsuperscript{st} Patricia Wollstadt}
\IEEEauthorblockA{\textit{Honda Research Institute Europe GmbH} \\
Offenbach/Main, Germany \\
patricia.wollstadt@honda-ri.de, 0000-0002-7105-5207}
\and
\IEEEauthorblockN{2\textsuperscript{nd} Sebastian Schmitt }
\IEEEauthorblockA{\textit{Honda Research Institute Europe GmbH} \\
Offenbach/Main, Germany \\
sebastian.schmitt@honda-ri.de, 0000-0001-7130-5483 } 
}

\maketitle

\begin{abstract}
An important issue during an engineering design process is to develop an understanding which design parameters have the most influence on the performance. Especially in the context of optimization approaches this knowledge is crucial in order to realize an efficient design process and achieve high-performing results. Information theory provides powerful tools to investigate these relationships because measures are model-free and thus also capture non-linear relationships, while requiring only minimal assumptions on the input data. We therefore propose to use recently introduced information-theoretic methods and estimation algorithms to find the most influential input parameters in optimization results. The proposed methods are in particular able to account for interactions between parameters, which are often neglected but may lead to redundant or synergistic contributions of multiple parameters. We demonstrate the application of these methods on optimization data from aerospace engineering, where we first identify the most relevant optimization parameters using a recently introduced information-theoretic feature-selection algorithm that accounts for interactions between parameters. Second, we use the novel partial information decomposition (PID) framework that allows to quantify redundant and synergistic contributions between selected parameters with respect to the optimization outcome to identify parameter interactions. We thus demonstrate the power of novel information-theoretic approaches in identifying relevant parameters in optimization runs and highlight how these methods avoid the selection of redundant parameters, while detecting interactions that result in synergistic contributions of multiple parameters.
\end{abstract}


\section{Introduction}

Optimizing the performance of systems of parts is a central task during an engineering design process. For example, in automotive or aerospace engineering, the shape of individual parts is commonly optimized to improve aerodynamic performance using computer aided design (CAD) methods. Typically, engineers wish to understand which changes in a shape, carried out during the optimization, lead to the improved behavior. Hereby, it is often of interest to account for interactions between parameters such as to identify which parameters influence a shape's fitness only when considered jointly \cite{Graening2009,Rath2011}. We therefore present a novel, information-theoretic approach for the identification of optimization parameters most relevant to changes in a shape's fitness, which accounts for interactions between parameters with respect to the fitness, such as to identify parameters that interact jointly with the target. We further utilize recently introduced information-theoretic measures to quantify interactions between features. We demonstrate the applicability of our approach on a set of realistic turbofan rotor blade optimization runs \cite{Kmec2018}, but strongly believe that it is of interest for a wide range of engineering design optimization scenarios.

Information theory \cite{Shannon1948} is a powerful tool for the analysis of dependencies between variables. Information-theoretic methods, such as the mutual information (MI), are model-free and are able to capture dependencies of arbitrary order, while requiring only minimal assumptions about the data for their estimation when using state-of-the-art estimators \cite{Kraskov2004}. These properties make information-theoretic measures particularly promising tools for the analysis of data in the engineering domain \cite{Graening2014}, for example, results from optimization runs \cite{Graening2012}. Here, the  relationship between parameters and the optimization objective is expected to be highly non-linear and the number of data samples is typically rather limited because the evaluation of fitness functions is costly. Furthermore, data distributions are typically not known and are expected to be highly biased due to the fact that data are generated by an optimization algorithm. As a result, high-quality global surrogate models that cover substantial parts of the search domain are most likely not available to understand optimization runs \cite{Kmec2018}. Thus, there is a need for methods that allow for a post-hoc analysis of optimization parameters and their influence on the optimization outcome.

We use a recently introduced algorithm for inferring relationships between variables that uses a conditional mutual information criterion (CMI) as a selection criterion \cite{Wollstadt2019,Novelli2019}. Using the CMI for selecting variables allows to account for interactions between variables such as redundancies, but also synergistic contributions \cite{Wollstadt2021}. Furthermore, we use the recently introduced partial information decomposition (PID) framework to investigate selected variables for interactions with respect to the target variable. We apply our approach to data from realistic turbofan blade aerodynamic optimization runs that use computational fluid dynamics (CFD) to evaluate a shape's fitness \cite{Kmec2018}. We propose a parametrization of the turbofan blade geometry that allows for application of the proposed algorithm and compare our algorithm's performance to related information-theoretic feature selection criteria. To our knowledge, this work is the first using PID for sensitivity analysis in aerodynamic optimization data.

\section{Methods}

\subsection{Optimization and Simulation Setup}

\begin{figure}[htbp]
    \centerline{
        \includegraphics[width=0.9\linewidth]{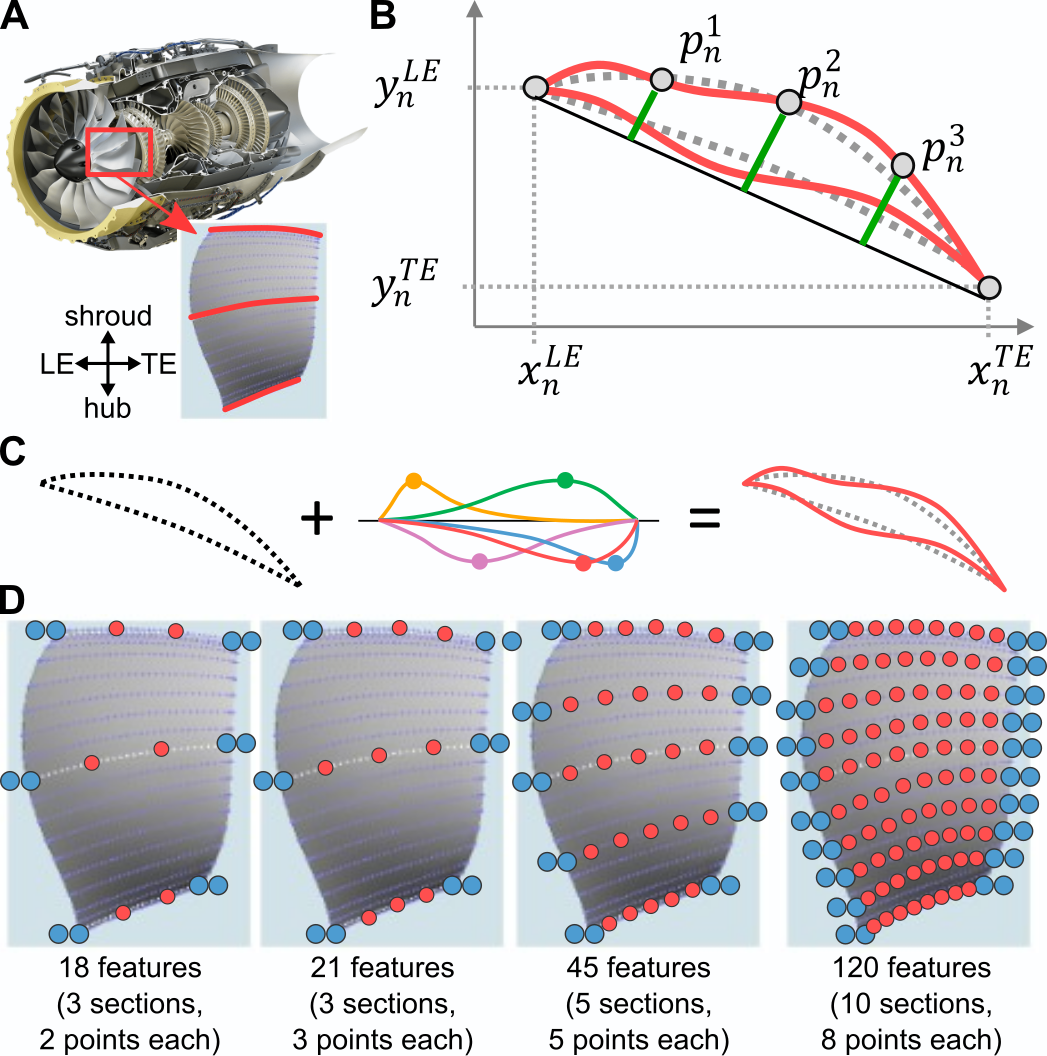}
    }
    \caption{Turbofan blade, modification and feature sets.
    \textbf{A} Investigated turbofan blade and Honda HF 120 jet engine, orange
    lines indicate cross-sections at which shape modifications were performed
    (LE: leading edge, TE: trailing edge).
    \textbf{B} Features selected form one blade section.
    \textbf{C} Modification of one blade section through addition of
    Hicks-Henne functions used during the shape optimization.
    \textbf{D} Location of the extracted feature sets, defined by varying numbers of
    sectional cuts through the geometry and number of points per section (red markers).
    Blue markers indicate leading and trailing edge features, each comprising two 
    features for the $x$- and $y$-coordinate of the edge, respectively.
    }
    \label{fig:turbofan_parametrization}
\end{figure}

We use data from realistic optimization runs on turbofan rotor blade geometries that
was previously described and  published in \cite{Kmec2018}. For details on the data generation 
process refer to the original publication. Fig.~\ref{fig:turbofan_parametrization}A 
shows a schematic of a turbojet engine and the corresponding turbofan rotor blade 
geometry. 

A rotor blade is optimized by starting with a baseline-shape that is
modified under the objective of minimizing a target function. The shape is 
modified by deforming three cross-sections of the shape, where each section is a
cylindrical cut of the geometry. We consider one section at the hub, 
one at mid-span height,  and one at the shroud of the blade as indicated by the red lines in the inset of Fig.~\ref{fig:turbofan_parametrization}A. 

Each section is deformed independently by the following manipulations:
(i) rotation of the section around the leading edge (LE) point, (ii) movement
of the section in the axial-meridional plane, and (iii) deformation of the
section profile by adding Hicks-Henne shape functions \cite{Hicks1978}, which
is a common approach in 2D airfoil design and is illustrated in
Fig.~\ref{fig:turbofan_parametrization}C. The Hicks-Henne function is defined as
\begin{equation}
    b(x, x_0) = \left[ \sin \left( \pi x^{\frac{\log(0.5)}{\log(x_0)}} \right) \right]^2,
    \label{eq:hicks_henne}
\end{equation}

\noindent where $x \in [0, 1]$ parametrizes the chord length of each section
and $x_0$ is the location of the maximum of each shape function. We placed the
maxima of $N_{HH}$ shape functions per section at equally spaced locations
along the cord length, $x_0(i) = \frac{i}{N_{HH}+1}$ where $i = 1,
\ldots,N_{HH}$. Considering all three possible manipulations, section rotation, movement, and
deformation with Hicks-Henne functions, the total number of free shape parameters 
is $N = 3\left(N_{HH} + 3\right)$.

For the optimization of shape parameters, we used a covariance matrix adaptation
evolutionary strategy (CMA-ES) \cite{Hansen2006} with a population size of
$\lambda = 12$ and $\mu = 4$ parents which we ran  for 161 generations, which amounts to 1932 evaluations, i.e.,\ data samples, per run. 
We used an initial step size of $\sigma =
0.05$ in relative units of the maximal allowed variation (i.e., a \SI{5}{\percent} initial variation).
We performed four optimization runs,
where two runs were performed with $N_{HH}=3$ and two runs with $N_{HH}=7$, which lead to $18$ and $30$ free parameters to be determined by the optimization, respectively. Each run was initialized using a different random seed. 
These parameter settings are derived from best practices which try to balance the exploration and exploitation capabilities of each optimization run, to arrive at manageable optimization run-times (each CFD simulation of a blade takes about \num{2} hours on \num{32} cores), and utilize the HPC infrastructure most efficiently. 

The optimization target was to maximize the aerodynamic efficiency of the rotor
blade at cruising conditions, which is estimated by calculating the isentropic efficiency of the blade,
\begin{equation}
    \eta =\frac{
        \left(\frac{P_{T, outlet}}{P_{T, inlet}}\right)^{\frac{\gamma-1}{\gamma}} - 1
    }{
        \frac{T_{T, outlet}}{T_{T, inlet}} - 1
    },
    \label{eq:sim_efficiency}
\end{equation}

\noindent where $P_T$ and $T_T$ are the mass-flow averaged total pressure and
total temperature at the specified location and $\gamma = 1.4$ is the heat
capacity ratio (see, for example, \cite{Baskharone2014}).

\begin{figure}[htbp]
    \centerline{
        \includegraphics[width=0.7\linewidth]{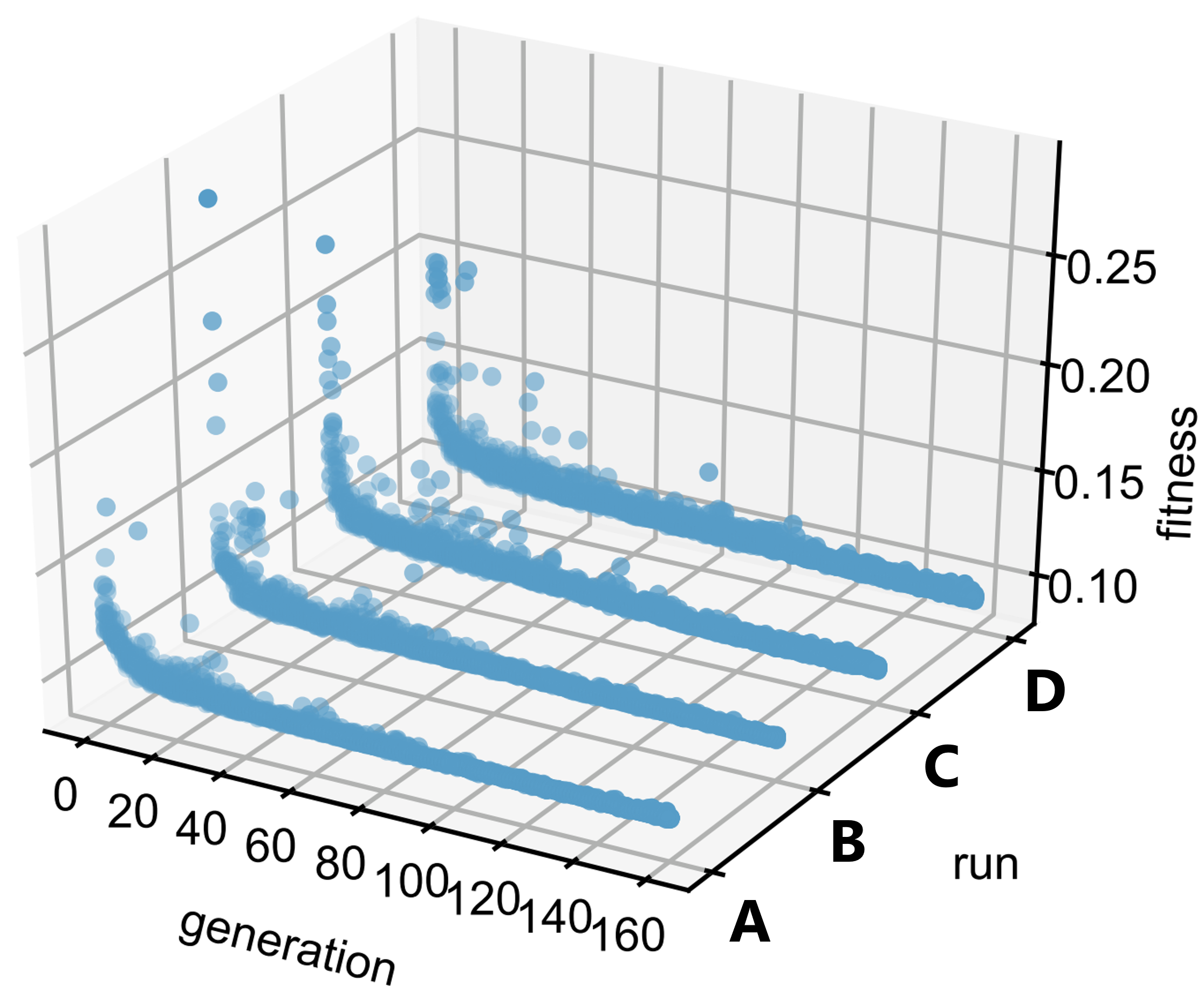} 
    }
    \caption{Raw fitness values over generations for four investigated optimizations.}
    \label{fig:raw_optimization_runs}
\end{figure}

The boundary conditions of the CFD simulation mimic the behavior of a jet engine under cruising conditions. 
Each blade was evaluated with a CFD simulation which employed the compressible flow solver \verb|steadyCompressibleMRFFoam| from the OpenFOAM CFD
suite (version \verb|foam-extend-3.2|), adapted to be more robust for trans-sonic simulations
\cite{Rusche2016}. The fitness of a blade was calculated as
\begin{equation}
    f = 1 - \eta_{avg} +  P,
    \label{eq:fitness}
\end{equation}

\noindent where $\eta_{avg}$ denotes the isentropic efficiency of the blade of Eq.~\eqref{eq:sim_efficiency}, averaged over the last 100 iterations of the solver. $P$ represents a penalty term that increases and thus worsens
the fitness if the CFD simulation does not show good convergence or if the generated blade
geometry is not feasible. 
See the original publication \cite{Kmec2018} for more details on the simulation setup, the optimization and the data generation.
The fitness values during the optimization runs as function of the generations is shown in Fig.~\ref{fig:raw_optimization_runs}.

\subsection{Feature Extraction of Turbofan Blade Geometries for Sensitivity Analysis}

We ran four optimizations with varying numbers of parameters for shape modification. 
In a next step, we wished to identify the locations at which modifications were most relevant to a blade's fitness. 
To apply the proposed information-theoretic approach, we first had to find suitable features that represented the blade geometry's surface and could be used as input features for the algorithm (e.g., \cite{Graening2008}).
To this end, we  considered multiple sectional cuts  through the blade geometry. 
At each sectional cut $n$, we placed $N_{points}$ points equally spaced along the chord line and recorded the absolute distance from the actual blade surface to the chord line at these locations (Fig.~\ref{fig:turbofan_parametrization}B). Furthermore, we considered the $x$- and $y$-coordinates of the leading edge (LE), $x_{n}^{LE}$ and $y_{n}^{LE}$, as well as the $x$- and $y$-coordinates of the trailing edge (TE), $x_{n}^{TE}$ and $y_{n}^{TE}$. We varied the number of points and sections used for the features to investigate the stability of results over various representations of the geometry.
We used 3 sectional cuts with 2 points, resulting in 18 features, 3  cuts with 3 points, resulting in 21 features, 5  cuts and 5 points resulting in 45 features, and 10  cuts and 8 points, resulting in 120 features  (Fig.~\ref{fig:turbofan_parametrization}D).

\subsection{Information-Theoretic Preliminaries}

Before introducing the algorithm used to identify the most relevant locations
of modification, we introduce the necessary information-theoretic preliminaries
(for a more detailed introduction see \cite{Cover2006}).

The algorithm uses a conditional mutual information (CMI) to quantify the influence
a single feature has on the fitness, in the context of further features.
The CMI is defined as

\begin{equation}
    I(X;Y|Z) = \sum_{x\in\mathcal{A}_X, y\in\mathcal{A}_Y, z\in\mathcal{A}_Z}
        p(x,y,z) \log \frac{p(x|y,z)}{p(x|z)},
    \label{eq:conditional_mutual_information}
\end{equation}

\noindent where $X$, $Y$, $Z$ are random variables with realizations $x$, $y$,
$z$, and $p(x)$ is a shorthand for the probability distribution $p(X=x)$. The
CMI quantifies the average information that $X$ has about $Y$, given the
outcome of $Z$ is known. The CMI is symmetric in $X$ and $Y$, and $I(X;Y|Z)
\geq 0$. Further, each random variable may also be replaced by a set of
variables, e.g., $\mathbf{X}$, and thus quantifying the information a set of
variables provides about a second variable, $Y$ or set of variables,
$\mathbf{Y}$.

Note that conditioning the information $X$ provides about $Y$ on a
third variable, $Z$, $I(X:Y|Z)$ may have two effects: first, information that is
\textit{redundantly} present in both $X$ and $Z$ about $Y$ is removed from the
information $X$ alone provides about $Y$ (as measured by the unconditioned
MI, $I(X;Y)$). Second, information that is provided
\textit{synergistically} by $X$ and $Z$ together about $Y$ is added to the
information $X$ alone is providing about $Y$ \cite{Williams2010}. Hence, the 
CMI quantifies the information $X$ provides \textit{uniquely} about $Y$ and the 
information $X$ and $Z$ provide jointly about $Y$ in a \textit{synergistic} fashion; 
at the same time, \textit{redundant} contributions in $X$ and $Z$ about $Y$ are 
excluded. See also \cite{Wollstadt2021} for a discussion of the use of the CMI for 
feature selection.

As an example of synergistic information contribution, consider a binary \texttt{xor}-gate 
with iid. inputs, $X$ and $Z$, and output $Y$. Inputs $X$ and $Z$ alone, each provide 
no information about the output $Y$, such that $I(X;Y)=I(Z;Y)=0$. Only by conditioning 
on the second input, the information the first input provides is \enquote{decoded} and 
$I(X;Y|Z)=I(Z;Y|X)=1$. Here, the two inputs provide information about the output in an 
exclusively synergistic fashion.
 
\begin{figure*}[htbp]
    \centerline{
        \includegraphics[width=0.9\textwidth]{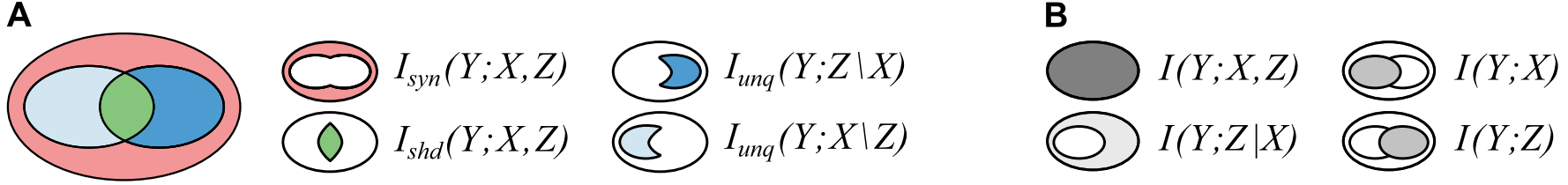}
    }
    \caption{
        \textbf{A} Partial information decomposition diagram: decomposition
        of the joint mutual information, $I(Y;X,Z)$ into unique information of
        each input variable (light and dark blue), redundant information
        (green), and synergistic information (red).
        \textbf{B} Corresponding, classical information-theoretic terms.
    }
    \label{fig:pid}
\end{figure*}

The framework to decompose the information two variables contribute about
a third into unique, redundant, and synergistic contributions has only recently 
been introduced and is termed \textit{Partial Information Decomposition} (PID) 
\cite{Williams2010} (Fig.~\ref{fig:pid}A, see also \cite{Gutknecht2020,Makkeh2021}). 
PID extends classical information theory by providing axioms 
that allow to decompose the joint information two input variables $X$ and $Z$ provide about 
a target variable $Y$, $I(Y;X,Z)$, into the information provided
uniquely by each $X$ and $Y$, information provided redundantly by $X$ and $Y$,
and information provided synergistically when considering $X$ and $Y$ jointly.
Note that such a detailed decomposition of the information contributed by two variables about a third was not possible using existing information-theoretic concepts, e.g., the (C)MI or Shannon entropy,  as shown by Williams and Beer \cite{Williams2010} and
illustrated in Fig.~\ref{fig:pid}B.

In the present work, we use the PID framework to identify interactions between 
features with respect to the blade's fitness. In particular, we estimate the 
synergistic information contribution of features and sets of features to identify 
those feature combinations that provide information about the fitness primarily 
when considered jointly.

\subsection{Identification of Most Relevant Features using
Information-Theoretic Feature Selection Algorithm}

We used a recently introduced forward-selection algorithm for feature selection
\cite{Wollstadt2019,Novelli2019,Wollstadt2021} to identify the most relevant blade
features with respect to the optimization outcome. The algorithm uses a CMI 
criterion for iterative feature selection, which measures the MI 
between a feature to be selected and the fitness, conditional on all already 
selected features. Thus, the CMI criterion, includes features not only based 
on their individual (unique) information contribution to the fitness, but also accounts for 
synergistic effects between the currently considered feature and the already selected 
feature set. Lastly, the inclusion criterion ensures that redundancies between 
features are avoided. For a detailed discussion of the algorithm and the CMI as 
a feature-selection criterion see \cite{Wollstadt2021}. See algorithm \ref{alg:feature_selection}.

\begin{algorithm}
    \caption{Forward feature selection}
    \label{alg:feature_selection}
    \begin{algorithmic}[1] 
        \Function{selectfeatures}{$\mathbf{X},Y, \alpha_{crit}$} 
            \State $\mathbf{S} \gets \emptyset$ \Comment{Initialization of feature set}
            \While{$\mathbf{X} \neq \emptyset$} \Comment{Find next candidate feature}
                \State $F \gets \max_{X \in \mathbf{X}} I(X; Y|\mathbf{S})$
                \State $\alpha \gets \text{permutationtest}(I(F; Y|\mathbf{S}))$
                \If{$\alpha < \alpha_{crit}$} \Comment{Contribution is significant}
                    \State $\mathbf{S} \gets \mathbf{S} \cup F$  \Comment{Add candidate to feature set}
                    \State $\mathbf{X} \gets \mathbf{X} \setminus F$ 
                \Else \Comment{Contribution is not significant}
                    \State \textbf{break} \Comment{Terminate inclusion}
                \EndIf
            \EndWhile
            \State \textbf{return} $\mathbf{S}$\Comment{Final feature set}
        \EndFunction
    \end{algorithmic}
\end{algorithm}

The algorithm starts with an empty feature set $\mathbf{S}=\emptyset$, the
set of all input variables, $\mathbf{X}_0=\mathbf{X}$, and the target variable
$Y$. Features are selected iteratively, where in each iteration, $i$, the 
algorithm selects the feature that maximizes the criterion,

\begin{equation}
    F_i = \max_{X \in \mathbf{X}_i} I(X; Y|\mathbf{S}_i),
    \label{eq:cmi_criterion}
\end{equation}

\noindent where $\mathbf{X}_i \subseteq \mathbf{X}$ denotes the remaining input
variables in iteration $i$, and $\mathbf{S}_i$ the set of already selected
features. The identified maximum contribution is tested for statistical
significance using non-parametric permutation testing and a testing scheme
that controls the family-wise error rate (see
\cite{Novelli2019} for a detailed description of the test). 
If the information contributed by $F_i$ as measured
by the CMI is statistically significant, $F_i$ is included in the set of
selected features, $\mathbf{S}_i$ and removed from the set of remaining 
variables, $\mathbf{X}_i$,
\begin{equation}
    \begin{aligned}
        \mathbf{S}_{i+1} &= \mathbf{S}_i \cup F_i \, , \\
        \mathbf{X}_{i+1} &= \mathbf{X}_i \setminus F_i\, .
    \end{aligned}
    \label{eq:inclusion_update}
\end{equation}

Note that statistical testing of the CMI estimate is necessary because while 
in theory the CMI is zero for (conditionally) independent variables, this may 
not be the case when estimating the CMI from finite data, due to the known 
bias of information-theoretic estimators (e.g.,
\cite{Paninski2003}). Instead, the test
evaluates whether the estimate significantly differs from the distribution of
estimates from permuted data and thus tests the Null-hypothesis of no
dependence between the feature and the target in the context of the already
selected feature set. The statistical test not only handles the estimation bias,
but also provides an automatic stopping criterion for feature selection, because
the algorithms stops if no remaining variable provides significant information 
about the target, given the already selected feature set. The number of features 
included in the selected feature set can indirectly be influenced by changing the 
critical alpha-level, $\alpha_{crit}$, of the statistical test, i.e., the threshold 
an individual test in iteration $i$ has to pass to allow for inclusion of candidate feature $F_i$. 
We here used $a_{crit}=0.05$, where lowering $\alpha_{crit}$ leads to a more strict criterion and 
thus to the selection of fewer features in general, and vice versa.

For practical estimation, we use an implementation of the algorithm as part
of the IDTxl python toolbox \cite{Wollstadt2019,Novelli2019,Wollstadt2021}, which uses
a k-nearest-neighbor-based estimator for MI and CMI estimation from continuous
data \cite{Kraskov2004}, which---while not being bias-free---has shown to provide 
the most favorable bias properties compared to other approaches 
\cite{Kraskov2004,Khan2007,Doquire2012}.

\subsection{Post-hoc Analysis of Feature Interactions by Estimating Synergistic
Information Contribution}

After selecting the most relevant geometric features for each optimization run 
using the presented forward-selection algorithm, we identify interactions between features with respect 
to the fitness by estimating the synergy between all pairs of selected features 
and the fitness. We use a PID estimator introduced in \cite{Makkeh2018}, also implemented in the IDTxl toolbox \cite{Wollstadt2019}.

\section{Results}

\subsection{Identified Features and Interactions Between Features}

\begin{figure}[htbp]
    \centerline{
        \includegraphics[width=0.9\linewidth]{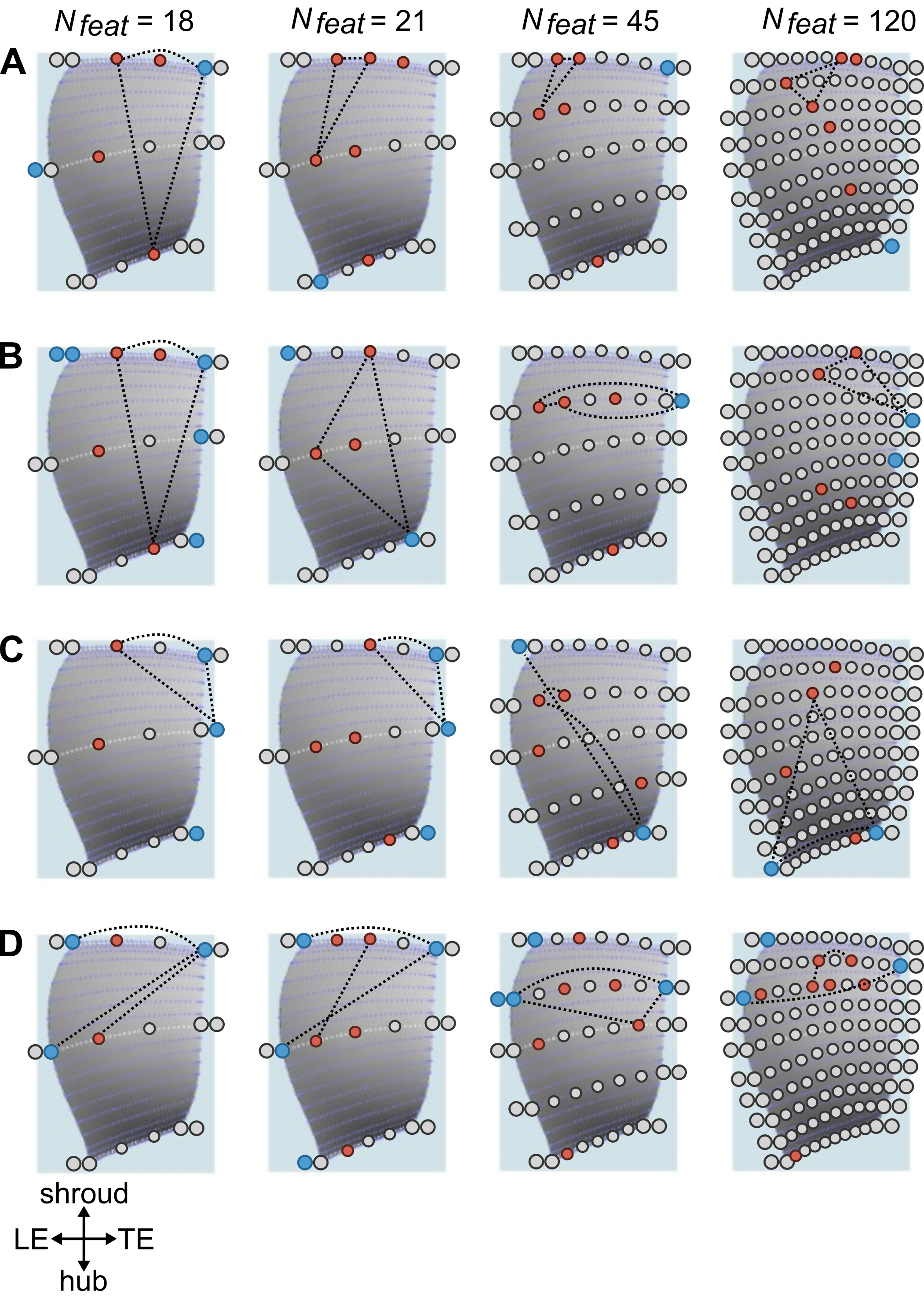}
    }
    \caption{Locations of selected features and identified interactions for four runs (rows A to D). Each column shows a different feature set, using 18, 21, 45, or 120 features respectively. Colored markers indicate selected features,  dotted lines indicate the three pairs with the highest interaction wrt.\ the blade's fitness as measured by the synergistic information. The meaning of the colors is the same as in Fig.~\ref{fig:turbofan_parametrization}D. }
    \label{fig:feature_location_interaction}
\end{figure}

The locations of features for the four optimization runs and the four extracted feature sets 
of the blade surface are shown in Fig.~\ref{fig:feature_location_interaction}. Here, the first two markers in each
row indicate the $x$- and $y$-coordinates of the leading edge,
$x_{n}^{LE}$, $y_{n}^{LE}$, while the last two markers indicate the coordinates of the trailing edge, $x_{n}^{TE}$, $y_{n}^{TE}$ (both are in blue).
The bottom row indicates the section closest to the  hub, while the top row indicates the section closest to the shroud. Markers between the first and last two markers in each row indicate geometric features from left to right, $p_{n}^{m}$, where $n \in \{1,\ldots, N\}$ indicates the section number from hub to shroud and $m \in \{1, \ldots, M\}$ indicates the feature index. 
Hence, the total number of input variables per feature set was $N_{feat} = NM + 4N$. Panels A and B, and panels C and D each show optimization runs with identical setup but different random initialization for $N_{HH}=3$ (A and B) and for $N_{HH}=7$ (C and D).  

Colored markers indicate relevant features identified by the algorithm. Dashed lines indicate the three pairs of features with highest synergy over all feature pairs.

We first note that the selected features are not completely consistent between runs which is expected. The data for each case was generated by an optimization run which is a highly structured process, and therefore,  the feature space is sampled very inhomogeneously. 
Additionally, the blade regions with the largest deformations differ between runs \cite{Kmec2018}, leading to  variations in the extracted features. However, there are regions which are identified to be important in all runs, for example, at around \SI{30}{\percent} chord length from the LE in the region from mid-span to blade tip (i.e.\ the upper forward region). This region is expected to have high influence due to the shock-system built-up \cite{Sonoda2013}. Similarly, the region near the TE, and in particular close to the tip,  is directly influencing the exit-flow angle and thus affecting the efficiency strongly. The location at the hub is also consistently identified as important, but the exact location along the chord line varies between the runs. 

Comparing the selected features of each run between the different feature sets  provides a consistent picture for the smaller feature sets $N_{feat}=18$, $21$, and  $45$.
The apparent differences can be understood by considering the peculiarities of the data and the PID-based selection method. 
First of all,  it is expected that for each feature set strong correlations and redundancy are present in the features, due to the deformation method used to generate the blades. 
Only three sections (at hub, mid-span and shroud) were allowed to change independently and the changes were linearly interpolated in-between, leading to many features being linear combinations of others. 
In addition, the Hicks-Henne-based deformations of each section also induces smooth changes with  possibly highly correlated and thus potentially redundant  neighboring features. 
Also, the optimization algorithm induces correlated changes of parameters, i.e.,\ blade regions,  once it starts to converge to some (local) optimum.
Therefore, the features  from the feature set with $N_{feat}=21$ selected on the mid-span section are  replaced by (Fig.~\ref{fig:feature_location_interaction}A and B) or augmented with (C and D)  more informative features on the second section from the tip. 
For  $N_{feat}=45$, high values of the redundancy are observed between the selected features and the not selected features which are close to the locations of the selected features form the smaller sets (not shown). 

For the largest feature set with $N_{feat}=120$ the selected features are consistent with the smaller feature sets in the above described manner for the case D, but  are only partially consistent or even seem inconsistent  for the other cases A, B and C.
This is understandable from the insights described above. Extracting 120 features from designs which are created with  only  18 (A and B) or 30 (C and D) independent parameters constitutes a vast over-parametrization of the independent influence factors, and results in huge redundancy in the feature set. In that case, the selected  features are  strongly influenced by the statistical variations of the rather few and highly structured 1932 data samples.
Multiple sets of features could  be selected which would be almost equally informative regarding the fitness,  but  which have different distribution of selected features over the blade region. Which set will be finally selected  is strongly influenced by its ability to describe the statistical fluctuations of the data set. From the theoretical perspective this is correct, as the selected features  represent the most informative features with respect to the fitness values \textit{for the given data set}. However, the value to the engineer might be limited,  as the most informative set does not necessarily represent the most important engineering design changes.

\subsection{Prediction of Optimization Results}
\label{sec:prediction}

To validate the identified set of relevant parameters for each combination of 
number of features and optimization run, we used the selected features to predict 
the fitness values of each blade across the optimization run. We compared the 
features selected by our algorithm to features selected by the FEAST toolbox 
\cite{Brown2012} and features selected by standard machine learning approaches (linear Pearson correlation, MI, decision trees, extra trees, random forest, 
LARS). 

The FEAST toolbox implements a variety of information-theoretic feature 
selection criteria based on the MI and applies them to rank features.
These criteria do not consider interactions between features, i.e., 
features are evaluated solely based on their \emph{individual}
contribution to the target. Hence, synergistic effects as well as redundancies are not 
accounted for (see also \cite{Wollstadt2021} for a comparison of the selection criteria
to the regular CMI). Also, the toolbox does neither provide means to handle estimator bias nor  
 an automatic stopping criterion for feature inclusion. As the toolbox only handles discrete variables, we binned the data prior to feature selection. 

We used the following selection criteria implemented in FEAST: 
Joint MI (JMI) \cite{Yang2000}, 
MI Maximization (MIM) \cite{Lewis1992}, 
Max-Relevance Min-Redundancy (MRMR) \cite{Peng2005}, 
Conditional MI Maximization (CMIM) \cite{Fleuret2004},
Double Input Symmetrical Relevance (DISR) \cite{Meyer2006}, 
Conditional Infomax Feature Extraction (CIFE) \cite{Lin2006},
Interaction Capping (ICAP) \cite{Jakulin2005}, 
Conditional Redundancy \cite{Brown2012},
Relief \cite{Kira1992}, 
and the CMI estimated from binned data.
We predicted the fitness from the different selected feature sets using 
$k$-nearest-neighbor regression with number of neighbors,
$k=1$.  Since the FEAST 
toolbox does not provide a stopping criterion, but just ranks the features by importance, 
we performed predictions from feature sets up to a size of 10 features, which was the maximum 
feature set size identified by our algorithm through statistical testing.

\begin{figure*}[htbp]
    \centerline{
        \includegraphics[width=0.85\linewidth]{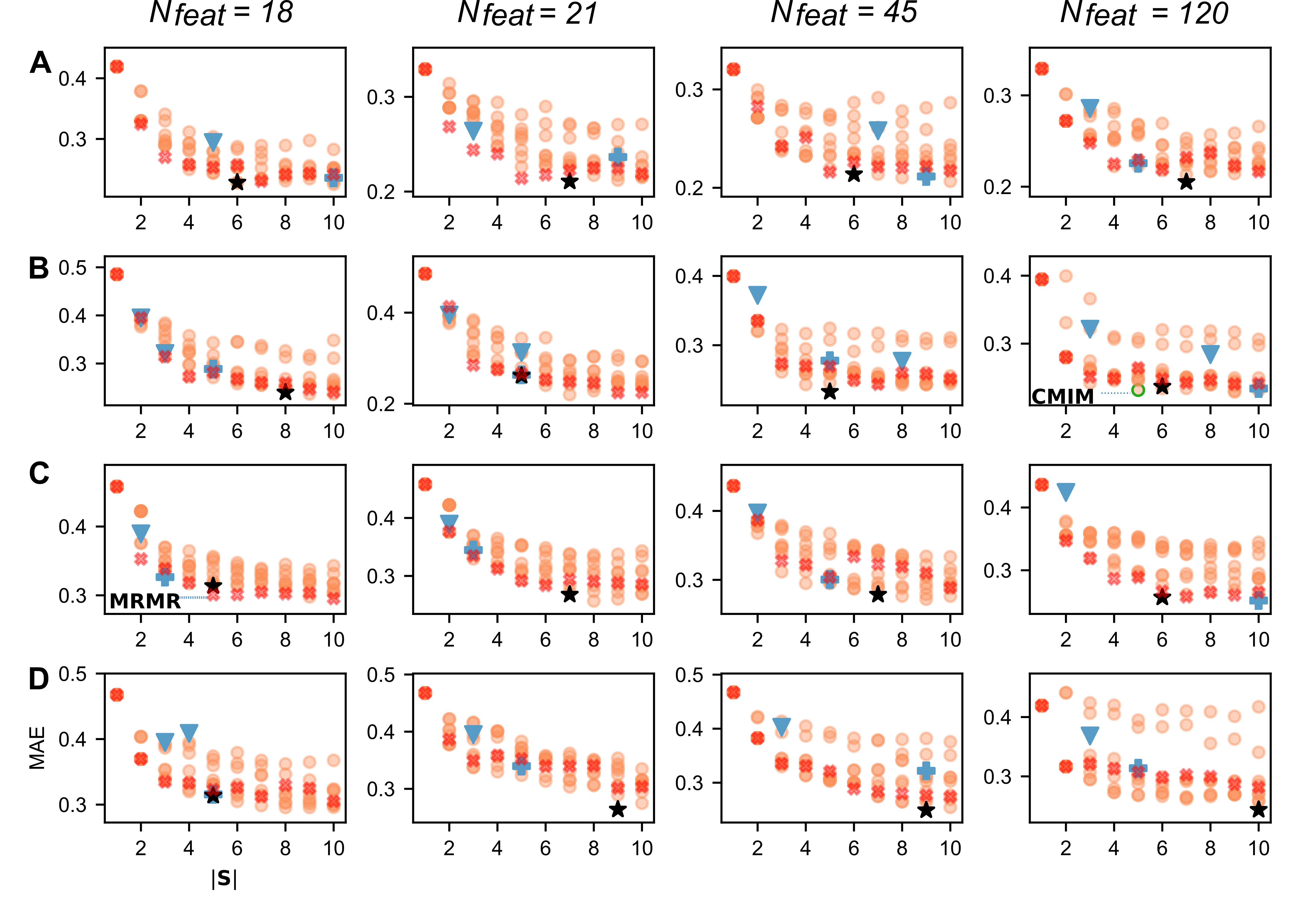}
    }
    \caption{Validation of feature set through $k$-nearest-neighbor-prediction of optimization outcome
    from selected feature sets (mean absolute error, MEA, y-axis). The x-axis indicates the size of the feature set, $|\mathbf{S}|$ (see main text). Each row shows results for one optimization run A-D. Each 
    column shows one feature set with $N_{feat}=18$, $21$, $45$, and $120$ features.
    Orange markers indicate prediction results for FEAST feature sets of different size, 
    blue markers indicate prediction results from machine learning approaches. Blue crosses are the feature sets selected with LARS while red crosses are those selected by MRMR.  
    The black star ($\bigstar$) indicates results from our algorithm. Annotations indicate the two cases where another method gave better prediction results than our method for a smaller or equally large feature set.}
    \label{fig:knn_prediction}
\end{figure*}

Prediction results for various identified feature sets using the FEAST toolbox, a selection of standard feature-selection
methods from machine learning, and our proposed algorithm are shown in Fig.~\ref{fig:knn_prediction}. The algorithms we compared our approach against, do not provide a stopping criterion, but only rank features by their importance. Hence, for each algorithm we predicted the fitness using various sets $\mathbf{S}$ of the highest-ranked variables to allow for comparison to our solution. The plots show the prediction error from various sets of sizes up to 10, i.e.,\ $|\mathbf{S}|=1,\dots,10$.
In many cases, the feature sets from standard machine learning approaches did not provide accurate predictions. Only for run C and D with $N_{feat}=18$, run A with $N_{feat}=45$ and for runs B and C with the largest feature set,  the predictions of one standard method allowed for rather accurate predictions compared to selected feature sets of the same size.  The features sets selected with the MRMR method, as one of the best performing methods from the FEAST toolbox, performed quite well, but only when a small number of features was selected and the relative performance dropped for larger sets of selected features. The proposed method based on CMI feature selection performed well for all studied situations as it consistently gave a good trade-off solution with respect to feature set size and prediction accuracy.   
In 14 out of the 16 considered runs and number of features, our algorithms selected
the best feature set among all feature sets of the same or smaller size. In 6
of these cases, the selected feature set performed best across  all feature sets of any size.
The other methods did not provide feature sets with such consistently good prediction performance, as can be seen, for example, for the LARS (blue crosses)  and MRMR (red crosses) method in Fig.~\ref{fig:knn_prediction}, which performed well for some configurations, but did not return good results consistently. 

Generally, we observed that many different feature sets led to similar prediction performance, especially for the largest feature set with $N_{feat}=120$, which supports our previous analysis that this parametrization lead to highly redundant and correlated features.
Nevertheless, the proposed CMI-based feature selection algorithm still managed to identify meaningful feature sets which were not too large and which allowed for good prediction performance.

\section{Discussion}

We applied a recently introduced information-theoretic approach to feature selection \cite{Wollstadt2021} in sensitivity analysis for optimization data. 
A strong conceptual and practical advantage of the proposed feature selection approach is its ability 
to account for interactions between variables when selecting features, such that the selection of redundant features is avoided while features that contribute information in a synergistic fashion together with other features are included. A further significant advantage of the used algorithm for the present application is the ability to automatically determine the number of relevant features by means of statistical testing, whereas for most established methods the number of features has to be fixed in advance. Furthermore, we used the recently introduced partial information decomposition (PID) framework \cite{Williams2010} to identify feature interactions.

We successfully applied the approach to four realistic aerodynamic optimization runs, where we showed that the feature sets identified by the proposed algorithm always provided a good trade-off solution with respect to feature set size and prediction performance. We showed that in most of the cases (14 out of 16) the selected feature set could be used to predict the optimization's objective function with smaller error than using feature sets of the same size or smaller identified through existing approaches. 

Central to the proposed approach is its ability to identify feature sets while accounting for interactions between features and to identify synergistic interactions. This property is especially desirable in application domains where optimization parameters are expected to show interactive effects on the target function. 
Such an analysis was previously not possible using the MI or its extensions, for example, the interaction information \cite{McGill1954}, which was proposed for the analysis of interactions in design data in earlier studies (e.g., \cite{Graening2009,Rath2011}). However, it was shown that these measures are not able to disentangle redundant and synergistic contributions and that such a contribution required the axiomatic extension of classical information theory as was done in the PID framework \cite{Williams2010}  (see also \cite{Gutknecht2020}). Accordingly, the development of information-theoretic filtering methods accounting for interactions has not advanced in recent years such that the methods employed here, which often assume variable independence, are still a common approach (e.g., MRMR \cite{Alnuaimi2020,Cai2018}). We believe that this stagnation is partially due to the inherent lack in classical information theory to describe multivariate information contributions that has only become available with the introduction of PID \cite{Williams2010,Wollstadt2021}.
Hence, PID enables the information-theoretic quantification of interactions in design applications as defined in \cite{Graening2014}: \enquote{a design interaction is defined as a unique dependency between design and objective parameters from which all dependencies of lower ordinality are removed}. 

The algorithm used for feature selection employs statistical testing to handle the bias in information-theoretic estimates. Statistical testing furthermore provides an automatic stopping criterion as it can reveal
that an estimate is not significantly different from an estimate from data with no relationship. Using statistical testing in feature selection has been proposed, for example, by \cite{Tsimpiris2012}. However, the approach used here is the first to rigorously control the family-wise error rate when testing repeatedly during iterative feature selection \cite{Novelli2019}.

The used algorithm accounts for redundant and synergistic contributions during the identification of relevant features by conditioning on the set of all already selected features. A limitation is here that due to the iterative inclusion, variables that provide purely synergistic information can not be detected. To handle this latter scenario, one may start feature selection with a non-empty set, e.g., some random subset or a subset informed by prior information. Alternatively, one may include variable tuples instead of individual variables \cite{Lizier2012b}. 

A further limiting factor is the number of features that the algorithm is able to select given a certain amount of data. If the selected feature set becomes too large, CMI-estimation suffers from the curse of dimensionality such that the CMI can no longer be estimated reliably from the available. As a result, the estimate fails to reach statistical significance and the algorithm terminates. However, in sensitivity analysis it is typically the goal to identify the set of \textit{most relevant} features that can still be meaningfully interpreted by a human. As shown here, the algorithm was able to identify up to 10 informative variables from less than 2000 highly biased samples.

Regarding the engineering task of identifying the most influential regions of the shape design the proposed approach gave satisfactory results, as  features located at known highly influential region were successfully identified. Also, the high degree of redundancy and correlations in the features sets, which is a natural consequence of the smoothness of the shape deformations, is handled well by the approach.

Future work may focus on a visualization and interpretation of the results to provide a more intuitive picture to the engineer who is potentially not vell-versed in information theory.

We conclude that the proposed algorithm \cite{Wollstadt2019,Novelli2019,Wollstadt2021}, together with the recently introduced PID framework \cite{Williams2010,Gutknecht2020,Makkeh2021} and suitable estimators \cite{Kraskov2004,Makkeh2018}, provides a valuable tool for the assessment of optimization outcomes in practical applications. In particular, the interaction-aware feature selection together with the estimation of synergistic effects allows to identify interactions between optimization parameters that was previously not possible using information-theoretic methods. Thus, the novel extension to information-theoretic analysis presented here provides powerful tools for quantifying relationships in a wide area of application domains that are concerned with the analysis of data from non-linear systems.



\end{document}